\documentclass[sigconf,natbib=true]{acmart}

\usepackage{subfigure}
 
\usepackage{amssymb}
\usepackage{multirow}
\usepackage{threeparttable}
\usepackage{makecell}
\usepackage{CJKutf8}
\usepackage{enumitem}
\usepackage{wrapfig,lipsum}

\newcommand\zemin[1]{\textcolor{blue}{#1}}

\newcommand\fw[1]{\textcolor{red}{#1}}

\AtBeginDocument{%
  \providecommand\BibTeX{{%
    \normalfont B\kern-0.5em{\scshape i\kern-0.25em b}\kern-0.8em\TeX}}}

\setcopyright{acmcopyright}
\copyrightyear{2022}
\acmYear{2022}
\acmDOI{10.1145/1122445.1122456}

\begin{document}
\title{Modeling Price Elasticity for Occupancy Prediction \\ in Hotel Dynamic Pricing}
\author{Fanwei Zhu}
\email{zhufanwei@zju.edu.cn}

\affiliation{%
  \institution{Zhejiang University City College}
  \city{Hangzhou}
  \country{China}
  \postcode{43017-6221}
}

\author{Wendong Xiao}
\email{xunxiao.xwd@alibaba-inc.com}

\affiliation{%
  \institution{Alibaba Group}
  \city{Hangzhou}
  \country{China}
  \postcode{43017-6221}
}

\author{Yao Yu}
\email{sichen.yy@alibaba-inc.com}

\affiliation{%
  \institution{Alibaba Group}
  \city{Hangzhou}
  \country{China}
  \postcode{43017-6221}
}

\author{Ziyi Wang}
\email{jianghu.wzy@alibaba-inc.com}

\affiliation{%
  \institution{Alibaba Group}
  \city{Hangzhou}
  \country{China}
  \postcode{43017-6221}
}

\author{Zulong Chen}
\email{zulong.czl@alibaba-inc.com}

\affiliation{%
  \institution{Alibaba Group}
  \city{Hangzhou}
  \country{China}
  \postcode{43017-6221}
}

\author{Quan  Lu}
\email{luquan.lq@alibaba-inc.com}

\affiliation{%
  \institution{Alibaba Group}
  \city{Hangzhou}
  \country{China}
  \postcode{43017-6221}
}

\author{Zemin Liu}
\email{zmliu@smu.edu.sg}

\affiliation{%
  \institution{Singapore Management University}
  \country{Singapore}
}

\author{Minghui Wu}
\email{mhwu@zucc.edu.cn}

\affiliation{%
  \institution{Zhejiang University City College}
  \city{Hangzhou}
  \country{China}
  \postcode{43017-6221}
}

\author{Shenghua  Ni}
\email{shenghua.nish@taobao.com }

\affiliation{%
  \institution{Alibaba Group}
  \city{Hangzhou}
  \country{China}
  \postcode{43017-6221}
}

\newcommand{\eg}{\textit{e.g.}}
\newcommand{\xeg}{\textit{E.g.}}
\newcommand{\ie}{\textit{i.e.}}
\newcommand{\xie}{\textit{I.e.}}
\newcommand{\etc}{\textit{etc.}}
\newcommand{\etal}{\textit{et al.}}
\newcommand{\wrt}{\textit{w.r.t}}
\newcommand{\mb}[1]{\mathbf{#1}}
\newcommand{\mc}[1]{\mathcal{#1}}
\newcommand{\figref}[1]{Fig.~\ref{#1}}
\newcommand{\tableref}[1]{Table~\ref{#1}}
\newcommand{\sectref}[1]{Sect.~\ref{#1}}
\newcommand{\B}[2]{\mathcal{B}_{#1}^{#2}} 
\newcommand{\xs}[2]{\mathcal{S}_{#1}^{#2}} 
\newcommand{\s}[2]{s_{#1}^{#2}} 
\newcommand{\st}[2]{\mathbf{\tau}_{#1}^{#2}} 
\newcommand{\x}[1]{\mathbf{x}_{#1}} 
\newcommand{\uv}[2]{\mathbf{v}_{#1}^{#2}} 
\newcommand{\p}[1]{\mathcal{P}_{#1}}
\newcommand{\f}[1]{\mathcal{F}_{#1}}
\newcommand{\iv}[1]{\mathbf{v}_{#1}}  
\newcommand{\fs}{g}   
\newcommand{\samp}{\mathbb{Y}}
\newcommand{\stitle}[1]{\vspace{2mm} \noindent {\bf #1}}
\newcommand{\sstitle}[1]{\vspace{1.5mm} \noindent {\emph{#1}}}
\newcommand{\sa}{SASNet}
\newcommand{\vp}[2]{\mathbf{c}_{#1}^{#2}} 

\begin{abstract}
Demand estimation plays an important role in dynamic pricing where the optimal price can be obtained via maximizing the revenue based on the demand curve.
In online hotel booking platform, the demand or occupancy of rooms varies across room-types and changes over time, and thus it is challenging to get an accurate occupancy estimate. 
In this paper, we propose a novel hotel demand function that explicitly models the price elasticity of demand for occupancy prediction, and design a price elasticity prediction model to learn the dynamic price elasticity coefficient from a variety of affecting factors. Our model is composed of carefully designed elasticity learning modules to alleviate the endogeneity problem, and trained in a multi-task framework to tackle the data sparseness.  
We conduct comprehensive experiments on real-world datasets and validate the superiority of our method over the state-of-the-art baselines for both occupancy prediction and dynamic pricing. 



\end{abstract}

\maketitle

\section{Introduction}
\label{sec:intro}
Dynamic pricing~\cite{dynamicPricing2015}, which determines the optimal prices of products or services based on the dynamic market demands, has received considerable attention in research~\cite{demand_elasticity2018,mannagementModel2006} and industries~\cite{hotel2004,dynamic_airline2019,comparison2003,customized2018,yang2014matching,roozbehani2010dynamic,deeplearningHotel2019}.
In online hotel reservation platforms (\eg, booking.com, Fliggy), dynamic pricing is extremely important as similar hotels on the platform compete to share the market demand, and the inventory (\ie, the available rooms) of each hotel is perishable on each day. Thus, a good pricing policy can benefit the matching of supply and demand, and improve the overall revenue. In practice, most pricing strategies recommend an optimal price to maximize the revenue based on a \textit{demand curve} \cite{dynamicPricing2015} that depicts the relationship between the price of a room and the demanded rooms, or particularly referred to as \emph{occupancy}, at that price. Therefore, occupancy estimation is the key to the success of dynamic pricing. 

However, an accurate occupancy estimate is challenging as the demand curve exhibits different patterns over time due to the diversity in hotel characteristics and external influences such as events, seasonality, \etc\
Regression models are widely adopted in existing works for occupancy prediction. For example, Ye \etal ~\cite{customized2018} use a Gradient Boosting Machine (GBM) ~\cite{friedman2001greedy} to map a set of raw features to an estimated booking probability for Airbnb's listings. Zhang \etal ~\cite{deeplearningHotel2019} capitalize on a seq2seq model \cite{sutskever2014sequence} to predict the future occupancy based on the hotel features and statistics. However, these regression models may suffer from the \textit{data sparseness} as many rooms only have reservations on certain days while the reservation prices are generally in a narrow range, and \textit{endogeneity problem}~\cite{econometrics2008} as many features are price dependent (\eg, the feature of historical sales is positively correlated with price). 
To address this, in this paper, we aim at a more accurate and explainable occupancy prediction approach for dynamic hotel pricing.

\stitle{Motivation.} Our idea is inspired by the concept of ``price elasticity" in economics~\cite{econometrics2008}. Particularly, we observe that though price is a significant affecting factor for hotel occupancy (\eg, the more a hotel lowers the price, the higher the occupancy rate), the sensitivity of occupancy to price change (or price elasticity of demand) could be quite different. 
\textit{On the one hand}, the price elasticity of hotels with different characteristics is essentially different, \eg, the occupancy of luxury hotels are less price-sensitive than budget hotels as they have less substitutes in the market place. 
\textit{On the other hand}, even for a same hotel room, its price elasticity could be varying across time due to the external factors such as seasonality and market trend. For example, hot spring hotels are less price-sensitive in winter as the demand is seasonal. 
\textit{Moreover}, the price elasticity of hotels in a shared market synchronously fluctuates due to the competition among them. For example, discount on a hotel room would drive higher demand for it, and at the same time lead to the lower occupancy for its competitors.

\stitle{Proposal.} Such observations motivate us to explicitly model the \textit{price elasticity} of demand in occupancy estimation. Specifically, we define the occupancy of a hotel room as a function $F(P,\beta)$ of its selling price $P$ and the price elasticity coefficient $\beta$, and propose a price elasticity model (PEM) to learn the dynamic price elasticity coefficient in the occupancy function. Once $\beta$ is inferred, the optimal price can be easily obtained based on the demand curve $F(P,\beta)$.

In addition to better modeling the demand dynamics, the elastic demand function brings two other advantages: 
1) the price elasticity of rooms are inherently similar to the price elasticity of hotels they belong to. Thus, the hotel-level reservation data can be used to handle the data sparsity issue and enhance the model training; 2) the features affecting $\beta$ are less dependent on price but more relevant to the room characteristics and external influences. Thus, it is promising to alleviate the endogeneity problem with proper model design.

Specifically, we comprehensively examine the features affecting the price elasticity and identify three orthogonal groups of factors: \textit{competitive factors} (\ie, the influence of competitive hotels in a same market), \textit{temporal factors} (\ie, the effect of hotel popularity and market trend), and \textit{characteristic factors} (\ie, the effect of basic features). We develop a competitiveness representation module and a multi-sequence fusion module to capture the competitive factors and the temporal factors, and attentively integrate them for occupancy prediction. Our model is trained in a multi-task learning framework using hotel occupancy prediction task to assist room occupancy prediction, to handle the data sparsity issue.

\stitle{Contributions.} Our contributions are three-fold. (1) We propose to utilize the price elasticity of demand to enhance the occupancy estimation and formally define a novel elastic demand function for hotel dynamic pricing. 
(2) We develop a multi-task price elasticity learning framework that tackles the data sparsity of room-level occupancy with the assistance of hotel-level occupancy prediction, and alleviate the endogeneity problem with carefully designed modules. 
(3) Extensive experiments on two real-world datasets validate the superiority of our method for both occupancy prediction and dynamic pricing.

\section{Approach}
\label{sec:model}
In this section, we define a novel demand function in hotel dynamic pricing, and present the proposed price elasticity learning model.
\subsection{Elastic Demand Function}
\label{subsec:problem}
In online hotel booking platform, a typical hotel provides multiple types of rooms for reservation on each day. The demand or occupancy of rooms varies across room-types (\eg, luxury rooms, economic rooms) and changes over time (\eg, weekday, weekend). 


To forecast the occupancy $O_r^t$ of a room type $r$ on a specific night $t$, as motivated in Sect.~\ref{sec:intro}, we propose to model $O_r^t$ as a function of the selling price $P_r^t$ and price elasticity coefficient $\beta_r^t$. We now concretely define the demand function based on two intuitions: 1) changes in $O_r^t$ are inversely related to changes in $P_r^t$ (\eg, demand usually falls when price rises); 2) how much $O_r^t$ changes is regulated by the price elasticity coefficient $\beta_r^t$.
Therefore, we can formalize the demand function in a concise yet effective way as
\begin{equation}
\vspace{-1mm}
    O_r^t= \bar{O}_r \cdot (P_r^t/\bar{P}_r)^{-\beta_r^t},
\label{eq:demand}
\end{equation}
where $\bar{O}_r$ and $\bar{P}_r$ are the constant benchmark occupancy and price, respectively, \eg, the average occupancy or price in one week.


Therefore, $\beta_r^t$ is the key to the success of occupancy prediction. Once $\beta_r^t$ is inferred, we can simply sample a set of prices $P_r^t$ in a reasonable range to calculate the corresponding $O_r^t$, and the optimal price is the one that maximizes the expected revenue
\begin{equation}
\vspace{-2mm}
     P_r^{t*} = \arg \max_{P_r^t} (P_r^t-P_r^0) \cdot O_r^t,
\end{equation}
\vspace{-1.5mm}
where $P_r^0$ is the cost price of room type $r$.

\subsection{Price Elasticity Prediction Model}
\label{subsec:PEM}
We now present the \textbf{P}rice \textbf{E}lasticity prediction \textbf{M}odel (PEM) to learn the price elasticity coefficient $\beta_r^t$ in Eq.~\eqref{eq:demand}. We identify three types of factors affecting the price elasticity in the hotel market:

(1) \emph{Competitive factors} such as the price and quality of competitors. For example, hotels that have better quality than its competitors are less price-elastic.
(2) \emph{Temporal factors} such as events, seasonality, popularity, \etc\ For example, the price elasticity of hot spring hotels are quite different in winter and summer due to the seasonal increase or decrease of its popularity. 
(3) \emph{Characteristic factors} such as hotel star, location and business district. Hotels with different characteristics naturally have different price elasticity. 

\begin{figure*}
    \centering
    \vspace{-2mm}
    \includegraphics[width=0.85\textwidth]{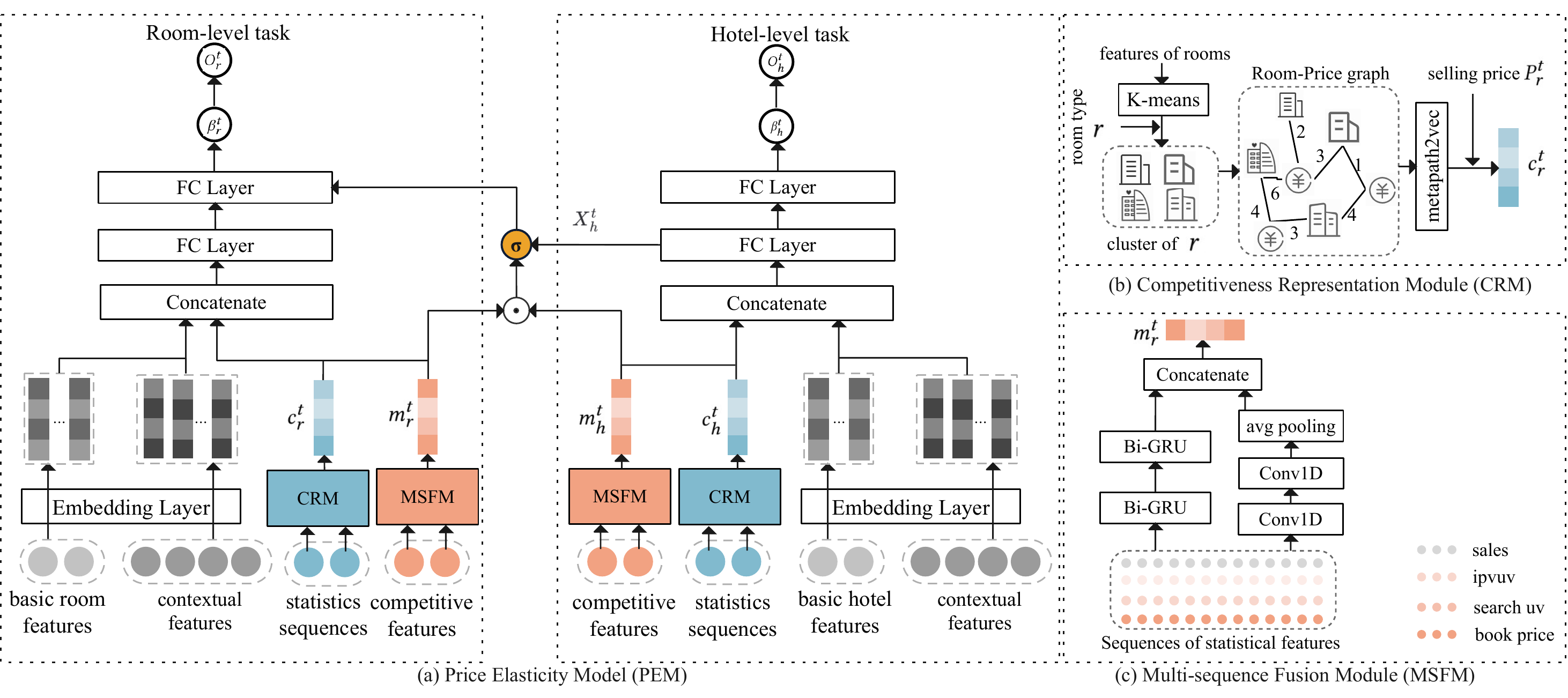}
    \vspace{-3mm}
    \caption{The proposed PEM model for price elasticity prediction.}
    \label{fig:archi}
    \vspace{-2mm}
\end{figure*}

Our PEM model is designed to leverage all these factors for dynamic demand learning. As shown in Fig.~\ref{fig:archi}, PEM is developed in a multi-task learning framework that simultaneously predicts the price elasticity coefficient $\beta_r^t$ of a room $r$ and the price elasticity coefficient $\beta_h^t$ of its belonging hotel $h$ at night $t$. For each prediction task, the input contains the aforementioned features, as well as the contextual features such as real-time price, holiday-or-not information. 
To alleviate the endogeneity problem, the basic features of location, hotel star \etc\ and contextual features are transformed into low-dimensional representations through an embedding layer, while the other features related to competitive factors and temporal factors are respectively processed with two sub-modules: Competitiveness Representation Module (CRM) and Multi-Sequence Fusion Module (MSFM), elaborated in the following subsections. Note that, we describe these modules in the context of room-level prediction task while for group-level task we only explain the specific inputs.

\subsubsection{Competitiveness Representation Module.} As discussed, the price elasticity of rooms in a shared market synchronously fluctuates due to the competition among them. 
To model this pattern, we propose the CRM module (Fig.~\ref{fig:archi}(b)) to learn the competitiveness of each selling price among a group of similar/competing rooms. The input of CRM consists of the basic features of all rooms $R$ (or hotels $H$), a specific room-type $r$  and the selling price $P_r^t$. 

To identify the competing groups, we cluster the rooms based on their basic features such as location, hotel star, and business district, with $K$-means clustering algorithm ($K$ equals to the number of business districts). The rooms in each cluster have similar qualities and usually compete with each other in the shared market, and thus the price change of one room would affect the others. 
For each group, we construct a heterogeneous room-price graph with two types of nodes, \ie, price and room, based on the reservation data in recent $30$ days. Each edge between a room node and a price node is weighted, indicating how many times the room was reserved at the connecting price. For each specific price, as its connections to rooms are sparse, we group the continuous price into several intervals (\eg, \$100-\$150, \$150-\$200) and represent each price interval with one single node.

From the constructed room-price graph, we conclude two intuitions:
1) Rooms that are frequently connected by the same price nodes are more similar in price-competitiveness
(\ie, they are better substitutes for each other);
2) Prices with more connections to the same rooms 
are more competitive than those have fewer connections.
Thus, we apply metapath2vec~\cite{metapath2vec2017} with metapaths \textit{room-price-room} and \textit{price-room-price}, to capture the above semantics and generate the embeddings for room and price nodes. 
%
Given the input selling price $P_r^t$, we can retrieve its competitiveness representation $\vp{r}{t}$ from the price embeddings, or average the representations of the two most similar price nodes if $P_r^t$ does not match any price node in the graph.
\subsubsection{Multi-sequence fusion module.}
The multi-sequence fusion model (Fig.~\ref{fig:archi}(c)) is designed to model the temporal factors that affect the price elasticity. The input of this module consists of the sequence of daily clicks (\ie, number of users who clicks the room), sequence of daily searches (\ie, number of searches for the room), and sequence of historical booking prices, of room-type $r$ in the latest $30$ days. For hotel-level task, the input statistics are the sum of all room statistics in a hotel.

As shown in Fig.~\ref{fig:archi}(c), the multiple sequences are fused with a two-layer Bi-GRU \cite{traffic2020} to model the combination of features and thus capture the popularity trends. 
We also apply two one-dimensional convolutions with different kernel sizes followed with an average pooling operation, to exploit the underlying local patterns for all the sequences.
The representations of popularity trends and local patterns are then concatenated as the final temporal vector ${m}_r^t$ .
\subsubsection{Multi-task losses.}
The competitiveness and temporal vectors are further employed to form a multi-task loss involving room-level and hotel-level price elasticity predictions for end-to-end training.

For room-level task, we concatenate both the competitiveness and temporal vectors $m_r^t||c_r^t$, as well as the contextual and basic room features, to form a unified feature representation, which is further fed into two fully-connected layers to estimate the price elasticity. We also leverage the intermediate representations of hotel-level features $X_h^t$ in room-level prediction by a gating mechanism based on the similarity of their popularity representations $m_r^t \cdot m_h^t$. The intuition is that the price elasticity of hotels will have more influence on rooms with similar popularity. 
For hotel-level task, $c_h^t$ and $m_h^t$ are concatenated and fed into an MLP to predict $\beta_h^t$. Finally, the estimated room occupancy $O_r^t$ and hotel occupancy $O_h^t$ can be derived by Eq.~\eqref{eq:demand}.

For model training, given $n$ selling price and occupancy tuples of historical room-level transactions $T_r=\{P_{ri}^t, O_{ri}^t\}_{i=1}^n$, 
and $m$ tuples of hotel-level transactions $T_h=\{P_{hj}^t, O_{hj}^t\}_{j=1}^m$ where $P_{h*}$ (or $O_{h*}$) is the sum of price (or occupancy) over all rooms in hotel $h$, we construct the overall loss by combining the room-level prediction loss $H_\delta(O_{ri}^t, \hat{O}_{ri}^t)$ and hotel-level prediction loss $H_\delta (O_{hj}^t, \hat{O}_{hj}^t)$:
\begin{equation}
\vspace{-2mm}
  \mathcal{L}=\lambda \sum_{i=1}^n H_\delta(O_{ri}^t, \hat{O}_{ri}^t) + (1-\lambda) \sum_{j=1}^m H_\delta (O_{hj}^t, \hat{O}_{hj}^t),
    \label{eq:loss}
\end{equation}
where $\lambda$ is a hyper-parameter to balance two tasks, and each prediction loss $H_\delta(y,\hat{y})$ is defined as the huber loss~\cite{Robust1964} with hyperparameter $\delta\in R^+$:
\begin{equation}
H_\delta (y,\hat{y})= \begin{cases}
\frac{1}{2} (y-\hat{y})^2, &  \text{if} \quad(y-\hat{y})\leq \delta; \\
\delta (y-\hat{y}) -\frac{1}{2}\delta^2, & \text{otherwise}.
\end{cases}
\end{equation}

\section{EXPERIMENTS}
\label{sec:expt}
In this section, we conduct extensive experiments on two real-world datasets to answer three questions: (\textbf{Q1}) How does the PEM model perform in occupancy prediction? (\textbf{Q2}) How does the proposed elastic demand function work in dynamic pricing? (\textbf{Q3}) How do different modules and losses contribute to the performance?

\subsection{Experimental Settings}
\subsubsection{Datasets.}
We construct two datasets by the hotel reservation logs collected from Fliggy\footnote{\url{www.fliggy.com}}, which is one of the most popular online travel platform in China.
In particular, \emph{Dataset-H} and \emph{Dataset-L} are based on the hotel reservation data in a high-booking season from 2020.12.17 to 2021.01.23 and a low-booking season from 2021.03.11 to 2022.04.17, respectively. 
In each dataset, the reservations in the first 30 days are employed as the training set to predict the occupancy and suggest price for the next 7 days. 
The details of the datasets are summarized in Table~\ref{tbl:datasets}.  
\begin{table}[h]
\center
\small
\caption{Summary of datasets.}
\vspace{-3mm}
\addtolength{\tabcolsep}{-0.6mm}
\resizebox{0.9\linewidth}{!}{
\begin{tabular}{@{}c|rrr|rrr@{}}
\toprule
\multicolumn{1}{c|}{\multirow{2}{*}{Datasets}} & \multicolumn{3}{c|}{Training}                                           & \multicolumn{3}{c}{Testing} \\
\cline{2-7}
\multicolumn{1}{c|}{}                                                                       & \# hotels & \# rooms & \# reservations &   \# hotels  & \# rooms & \# reservations \\
\cline{1-7}
\begin{tabular}[c]{@{}l@{}}Dataset-L \end{tabular}                                         & 12,727 &  16,676   & 712K &      7,859   &  10,023       &      222K        \\
   \begin{tabular}[c]{@{}l@{}}Dataset-H \end{tabular}                                     & 11,556 &   14,801 & 834K    & 9,533        & 12,804        &     169K             \\
\bottomrule
\end{tabular}
}
\vspace{-5mm}
\label{tbl:datasets}
\end{table}


\subsubsection{Baselines.}
We compare our method with two groups of baselines, \ie, occupancy prediction and dynamic pricing.


\stitle{Occupancy prediction approaches.}
(1) \emph{GBDT} \cite{stochastic1999} is an additive regression model that gives the prediction in the form of an ensemble of decision trees. 
(2) \emph{Seasonal Autoregressive Integrated Moving Average (SARIMA)} \cite{warren2017occupancy}  combines autoregressive model with moving average to model time series exhibiting seasonality.
(3) \emph{DeepFM+Seq2seq} \cite{deeplearningHotel2019} is a sequence learning model for occupancy prediction which integrates DeepFM and Seq2Seq model to learn the interaction among different features in the sequence and regress the future occupancy.
(4) \emph{DeepAR} \cite{deepAR2019} is a probabilistic time series forecasting method based on auto-regressive recurrent network.
(5) \emph{TFT} \cite{temporal2020} combines recurrent layers with self-attention layers \cite{vaswani2017attention} to learn temporal relationships at different scales.


\stitle{Dynamic pricing approaches.}
(1) \emph{DNN-based pricing} \cite{deeplearningHotel2019} adopts a DNN model to regress the suggested price for a certain room at a specific date from the input features.
(2) \emph{Airbnb pricing} \cite{customized2018} uses a customized regression model to suggest the price \wrt\ the booking probability for Airbnb listing-nights and applies personalized logic to optimize the suggestion.

Among these baselines, DeepFM+Seq2Seq and TFT are the state-of-the-art methods for occupancy prediction and general time series forecasting respectively. DNN-based pricing and Airbnb pricing are the only published work in dynamic hotel pricing.

\subsubsection{Implementation details.}
Our proposed PEM contains two fully-connected layers with 256 and 128 units, respectively. The embedding size of all features is set to 16. 
The multi-sequence fusion model contains three convolution kernels (sizes: 64, 128, 64) and a two-layer Bi-GRU module (number of units: 64, 32). {For training, the batch size is set to 512,} dropout rate is 0.2, $\delta=1.0$ and $\lambda=0.9$. 
For GBDT, we use the algorithm implemented by eXtreme Gradient Boosting (XGBoost) \cite{chen2015xgboost} where the number of trees and the max tree length in GBDT is set to 500 and 6 respectively; For SARIMA, we adopt the x13-auto-arima version \cite{gomez2000automatic} of the algorithm. The neural network used in DeepAR and TFT is one-layer LSTM \cite{hochreiter1997long} with 40 and 16 hidden cells respectively, number of heads in the multi-head attention layer of TFT is set to 2. The regression module of Airbnb is a three-layer MLP (number of hidden nodes: $128,32,1$). The other parameters are set as suggested in the original papers. 







\subsection{Performance Comparison}
\subsubsection{Evaluation metrics.} 
We adopt two common metrics MAPE and WMAPE \cite{de2016mean,tofallis2015better} for the evaluation of occupancy prediction.
For the evaluation of pricing strategy, we adopt the same metrics employed in hotel dynamic pricing \cite{customized2018,deeplearningHotel2019}: PDR, PDP, PIR, PIP and BR. The detailed description of these metrics can be referred to \cite{deeplearningHotel2019}.
Generally, BR measures the closeness between the suggested prices and the booked prices, and the other four metrics measures the profit improvement at the suggested prices. In practice, there should be trade-offs for these metrics to increase the overall revenue of the platform and to maintain business trust from hoteliers.


\begin{table}[]
 \caption{Evaluation on occupancy prediction (in percent).}
    \centering
    \vspace{-2mm}
    \resizebox{0.86\linewidth}{!}{
    \begin{tabular}{c|c|c|c|c}
    \toprule
   \multirow{2}{*}{Method} & \multicolumn{2}{c|}{Dataset-H} & \multicolumn{2}{c}{Dataset-L} \\
   \cline{2-5}
           &MAPE $\downarrow$ & WMAPE $\downarrow$  &MAPE $\downarrow$ & WMAPE $\downarrow$  \\
         \hline
        
          GBDT & 43.83 & 42.57 & 45.80 &46.51\\
          SARIMA & 28.56 & 40.03 & 30.51 & 41.10\\
          DeepFM+Seq2seq & 28.72 & 39.75 & 30.22& 41.13\\
          DeepAR & 28.11 & 39.78 & 30.24 & 40.86\\
          TFT & 27.96 & 39.22 & 30.03 & 40.37\\
          PEM & \textbf{27.88} & \textbf{39.03} & \textbf{29.86} & \textbf{40.21}\\
         \bottomrule
    \end{tabular}
   }
   \vspace{-2mm}
    \label{tbl:occupancy}
\end{table}


\begin{table}[htbp]
  \caption{Evaluation on pricing strategy (in percent).}
  \vspace{-1mm}
  \label{tbl:pricing}
  \resizebox{\linewidth}{!}{
  \begin{tabular}{c|c|c|c|c|c|c|c|c|c|c}
    \toprule
   \multirow{2}{*}{Method} & \multicolumn{5}{c|}{Dataset-H} & \multicolumn{5}{c}{Dataset-L} \\
   \cline{2-11}
           &PDR $\uparrow$ & PDP $\uparrow$  &PIR $\uparrow$ & PIP $\uparrow$ & BR $\downarrow$ & PDR $\uparrow$ & PDP $\uparrow$  & PIR $\uparrow$ & PIP $\uparrow$ & BR $\downarrow$ \\
         \hline
          DNN &54.0& 53.7&27.8 &28.1 & \ \ \textbf{7.8} &60.6 &59.4 &35.9 &37.0& \ \ 5.6  \\
          Airbnb &54.8 & 50.3 & 27.0 & 27.8 & \ \ {8.1} & 60.3 & 59.7& 34.3 &35.8 & \ \ \textbf{5.5} \\
          PEM & \textbf{56.2}& \textbf{55.5} &\textbf{30.1} &\textbf{30.8} & \ \ 8.5 & \textbf{61.3} & \textbf{60.5} &\textbf{37.9} &\textbf{38.8} & \ \ 6.0 \\
         \bottomrule
    \end{tabular}
}
\vspace{-5mm}
\end{table}

\subsubsection{Results and analysis.}
The comparison of occupancy prediction and dynamic pricing are shown in Tables~\ref{tbl:occupancy} and \ref{tbl:pricing}, respectively.
For occupancy prediction, PEM achieves the best performance in terms of both metrics, which validates the effectiveness of our elastic demand function and the price elasticity prediction model for occupancy forecasting (\textbf{Q1}).
For dynamic pricing, we observe:
1) PEM significantly outperforms DNN and Airbnb on the first four metrics related to revenue improvement. 2) The BR value of PEM is approaching DNN and Airbnb which are designed to directly regress the booking prices. However, for these baselines, the improvement in BR is obtained at the expense of much worse profit gains. In contrast, our PEM model achieves the best trade-offs among these metrics, {demonstrating} the effectiveness of our pricing strategy with elastic demand function (\textbf{Q2}).

We also deploy our model in the live environment of Fliggy hotel booking system and evaluate the platform revenue with Gross Merchandise Volumn (GMV)~\cite{miao2020data}, a metric commonly-used in e-commerce industry. In the online test for two weeks, the PEM model achieves an average of $7.42\%$ improvement in daily GMV compared with a baseline manual pricing strategy. The details of online experiments are omitted due to the space limitation.


\subsection{Ablation Study \textbf{(Q3)}}

We conduct an ablation study to investigate the contribution of each key module and loss. We compare PEM with three variants: 1) PEM-C: without competitiveness representation module; 2) PEM-M: without multi-sequence fusion module; 3) PEM-L: without hotel-level occupancy prediction task.
The experimental results illustrated in Table~\ref{tbl:ablation} verify that all the components and losses are essential, among which the multi-sequence fusion module contributes the most. 
Specifically, on Dataset-H, the accuracy of PEM-C, PEM-M and PEM-L in WMAPE  drops by $2.4\%$, $2.7\%$, $1.3\%$  respectively.


\begin{table}[htbp]
    \centering
     \caption{Ablation study}
    \begin{tabular}{c|c|c}
    \toprule
        Method & MAPE $\downarrow$ & WMAPE $\downarrow$ \\ \hline
        PEM-C & 29.01 & 40.15 \\ 
        PEM-M & 48.18 & 52.32 \\ 
        PEM-L & 28.39 & 39.81 \\ 
        PEM & \textbf{27.88} & \textbf{39.03} \\ 
        \bottomrule
    \end{tabular}
    \label{tbl:ablation}
\end{table}

\section{CONCLUSIONS}
\label{conclusion}
In this paper, we propose a novel elastic demand function that captures the price elasticity of demand in hotel occupancy prediction. We develop a price elasticity prediction model (PEM) with a competitive representation module and a multi-sequence fusion model to learn the dynamic price elasticity from a complex set of affecting factors. Moreover, a multi-task framework consisting of room- and hotel-level occupancy prediction tasks is introduced to PEM to alleviate the data sparsity issue. Extensive experiments on real-world datasets show that PEM outperforms other state-of-the-art methods for both occupancy prediction and dynamic pricing. PEM model has been successfully deployed at Fliggy and shown good performance in online hotel booking services.

\clearpage

\bibliographystyle{ACM-Reference-Format}
\bibliography{sample-base}


\begin{thebibliography}{27}


\ifx \showCODEN    \undefined \def \showCODEN     #1{\unskip}     \fi
\ifx \showDOI      \undefined \def \showDOI       #1{#1}\fi
\ifx \showISBNx    \undefined \def \showISBNx     #1{\unskip}     \fi
\ifx \showISBNxiii \undefined \def \showISBNxiii  #1{\unskip}     \fi
\ifx \showISSN     \undefined \def \showISSN      #1{\unskip}     \fi
\ifx \showLCCN     \undefined \def \showLCCN      #1{\unskip}     \fi
\ifx \shownote     \undefined \def \shownote      #1{#1}          \fi
\ifx \showarticletitle \undefined \def \showarticletitle #1{#1}   \fi
\ifx \showURL      \undefined \def \showURL       {\relax}        \fi
\providecommand\bibfield[2]{#2}
\providecommand\bibinfo[2]{#2}
\providecommand\natexlab[1]{#1}
\providecommand\showeprint[2][]{arXiv:#2}

\bibitem[Abbas and Zellner(2018)]%
        {demand_elasticity2018}
\bibfield{author}{\bibinfo{person}{Ali Abbas} {and} \bibinfo{person}{Maximilian
  Zellner}.} \bibinfo{year}{2018}\natexlab{}.
\newblock \showarticletitle{The effects of demand elasticity and selling price
  decisions on the value of information}. In \bibinfo{booktitle}{\emph{2018
  Annual IEEE International Systems Conference (SysCon)}}.
  \bibinfo{pages}{1--8}.
\newblock
\urldef\tempurl%
\url{https://doi.org/10.1109/SYSCON.2018.8369489}
\showDOI{\tempurl}


\bibitem[Chen et~al\mbox{.}(2015)]%
        {chen2015xgboost}
\bibfield{author}{\bibinfo{person}{Tianqi Chen}, \bibinfo{person}{Tong He},
  \bibinfo{person}{Michael Benesty}, \bibinfo{person}{Vadim Khotilovich},
  \bibinfo{person}{Yuan Tang}, \bibinfo{person}{Hyunsu Cho},
  \bibinfo{person}{Kailong Chen}, {et~al\mbox{.}}}
  \bibinfo{year}{2015}\natexlab{}.
\newblock \showarticletitle{Xgboost: extreme gradient boosting}.
\newblock \bibinfo{journal}{\emph{R package version 0.4-2}}
  \bibinfo{volume}{1}, \bibinfo{number}{4} (\bibinfo{year}{2015}),
  \bibinfo{pages}{1--4}.
\newblock


\bibitem[Choi and Mattila(2004)]%
        {hotel2004}
\bibfield{author}{\bibinfo{person}{S. Choi} {and} \bibinfo{person}{A.~S.
  Mattila}.} \bibinfo{year}{2004}\natexlab{}.
\newblock \showarticletitle{Hotel revenue management and its impact on
  customers’ perceptions of fairness}.
\newblock \bibinfo{journal}{\emph{Revenue and Pricing Management}}
  \bibinfo{volume}{2} (\bibinfo{year}{2004}), \bibinfo{pages}{303}.
\newblock
\urldef\tempurl%
\url{https://doi.org/10.1057/palgrave.rpm.5170079}
\showURL{%
\tempurl}


\bibitem[De~Myttenaere et~al\mbox{.}(2016)]%
        {de2016mean}
\bibfield{author}{\bibinfo{person}{Arnaud De~Myttenaere},
  \bibinfo{person}{Boris Golden}, \bibinfo{person}{B{\'e}n{\'e}dicte Le~Grand},
  {and} \bibinfo{person}{Fabrice Rossi}.} \bibinfo{year}{2016}\natexlab{}.
\newblock \showarticletitle{Mean absolute percentage error for regression
  models}.
\newblock \bibinfo{journal}{\emph{Neurocomputing}}  \bibinfo{volume}{192}
  (\bibinfo{year}{2016}), \bibinfo{pages}{38--48}.
\newblock


\bibitem[{den Boer}(2015)]%
        {dynamicPricing2015}
\bibfield{author}{\bibinfo{person}{Arnoud~V. {den Boer}}.}
  \bibinfo{year}{2015}\natexlab{}.
\newblock \showarticletitle{Dynamic pricing and learning: Historical origins,
  current research, and new directions}.
\newblock \bibinfo{journal}{\emph{Surveys in Operations Research and Management
  Science}} \bibinfo{volume}{20}, \bibinfo{number}{1} (\bibinfo{year}{2015}),
  \bibinfo{pages}{1--18}.
\newblock
\showISSN{1876-7354}
\urldef\tempurl%
\url{https://doi.org/10.1016/j.sorms.2015.03.001}
\showDOI{\tempurl}


\bibitem[Dong et~al\mbox{.}(2017)]%
        {metapath2vec2017}
\bibfield{author}{\bibinfo{person}{Yuxiao Dong}, \bibinfo{person}{Nitesh~V.
  Chawla}, {and} \bibinfo{person}{Ananthram Swami}.}
  \bibinfo{year}{2017}\natexlab{}.
\newblock \showarticletitle{metapath2vec: Scalable Representation Learning for
  Heterogeneous Networks}. In \bibinfo{booktitle}{\emph{Proceedings of the 23rd
  {ACM} {SIGKDD} International Conference on Knowledge Discovery and Data
  Mining, Halifax, NS, Canada, August 13 - 17, 2017}}.
  \bibinfo{publisher}{{ACM}}, \bibinfo{pages}{135--144}.
\newblock
\urldef\tempurl%
\url{https://doi.org/10.1145/3097983.3098036}
\showDOI{\tempurl}


\bibitem[Emeksiz et~al\mbox{.}(2006)]%
        {mannagementModel2006}
\bibfield{author}{\bibinfo{person}{Murat Emeksiz}, \bibinfo{person}{Dogan
  Gursoy}, {and} \bibinfo{person}{Orhan Icoz}.}
  \bibinfo{year}{2006}\natexlab{}.
\newblock \showarticletitle{A yield management model for five-star hotels:
  Computerized and non-computerized implementation}.
\newblock \bibinfo{journal}{\emph{International Journal of Hospitality
  Management}} \bibinfo{volume}{25}, \bibinfo{number}{4}
  (\bibinfo{year}{2006}), \bibinfo{pages}{536--551}.
\newblock
\showISSN{0278-4319}
\urldef\tempurl%
\url{https://doi.org/10.1016/j.ijhm.2005.03.003}
\showDOI{\tempurl}


\bibitem[Friedman(1999)]%
        {stochastic1999}
\bibfield{author}{\bibinfo{person}{J.H. Friedman}.}
  \bibinfo{year}{1999}\natexlab{}.
\newblock \bibinfo{booktitle}{\emph{Stochastic Gradient Boosting}}.
\newblock \bibinfo{type}{{T}echnical {R}eport}. \bibinfo{institution}{Stanford
  University}.
\newblock


\bibitem[Friedman(2001)]%
        {friedman2001greedy}
\bibfield{author}{\bibinfo{person}{Jerome~H Friedman}.}
  \bibinfo{year}{2001}\natexlab{}.
\newblock \showarticletitle{Greedy function approximation: a gradient boosting
  machine}.
\newblock \bibinfo{journal}{\emph{Annals of statistics}}
  (\bibinfo{year}{2001}), \bibinfo{pages}{1189--1232}.
\newblock


\bibitem[Gomez and Maravall(2000)]%
        {gomez2000automatic}
\bibfield{author}{\bibinfo{person}{Victor Gomez} {and} \bibinfo{person}{Agustin
  Maravall}.} \bibinfo{year}{2000}\natexlab{}.
\newblock \showarticletitle{Automatic modeling methods for univariate series}.
\newblock \bibinfo{journal}{\emph{A course in time series analysis}}
  (\bibinfo{year}{2000}), \bibinfo{pages}{171--201}.
\newblock


\bibitem[Hochreiter and Schmidhuber(1997)]%
        {hochreiter1997long}
\bibfield{author}{\bibinfo{person}{Sepp Hochreiter} {and}
  \bibinfo{person}{J{\"u}rgen Schmidhuber}.} \bibinfo{year}{1997}\natexlab{}.
\newblock \showarticletitle{Long short-term memory}.
\newblock \bibinfo{journal}{\emph{Neural computation}} \bibinfo{volume}{9},
  \bibinfo{number}{8} (\bibinfo{year}{1997}), \bibinfo{pages}{1735--1780}.
\newblock


\bibitem[Huber(1964)]%
        {Robust1964}
\bibfield{author}{\bibinfo{person}{Peter~J. Huber}.}
  \bibinfo{year}{1964}\natexlab{}.
\newblock \showarticletitle{{Robust Estimation of a Location Parameter}}.
\newblock \bibinfo{journal}{\emph{The Annals of Mathematical Statistics}}
  \bibinfo{volume}{35}, \bibinfo{number}{1} (\bibinfo{year}{1964}),
  \bibinfo{pages}{73 -- 101}.
\newblock
\urldef\tempurl%
\url{https://doi.org/10.1214/aoms/1177703732}
\showDOI{\tempurl}


\bibitem[Lim et~al\mbox{.}(2020)]%
        {temporal2020}
\bibfield{author}{\bibinfo{person}{Bryan Lim}, \bibinfo{person}{Sercan~O.
  Arik}, \bibinfo{person}{Nicolas Loeff}, {and} \bibinfo{person}{Tomas
  Pfister}.} \bibinfo{year}{2020}\natexlab{}.
\newblock \bibinfo{title}{Temporal Fusion Transformers for Interpretable
  Multi-horizon Time Series Forecasting}.
\newblock
\newblock
\showeprint[arxiv]{1912.09363}~[stat.ML]


\bibitem[Miao(2020)]%
        {miao2020data}
\bibfield{author}{\bibinfo{person}{Sentao Miao}.}
  \bibinfo{year}{2020}\natexlab{}.
\newblock \emph{\bibinfo{title}{Data-Driven Optimization in Revenue Management:
  Pricing, Assortment Planning, and Demand Learning}}.
\newblock \bibinfo{thesistype}{Ph.\,D. Dissertation}.
\newblock


\bibitem[Roozbehani et~al\mbox{.}(2010)]%
        {roozbehani2010dynamic}
\bibfield{author}{\bibinfo{person}{Mardavij Roozbehani},
  \bibinfo{person}{Munther Dahleh}, {and} \bibinfo{person}{Sanjoy Mitter}.}
  \bibinfo{year}{2010}\natexlab{}.
\newblock \showarticletitle{Dynamic pricing and stabilization of supply and
  demand in modern electric power grids}. In \bibinfo{booktitle}{\emph{2010
  First IEEE International Conference on Smart Grid Communications}}. IEEE,
  \bibinfo{pages}{543--548}.
\newblock


\bibitem[Salinas et~al\mbox{.}(2019)]%
        {deepAR2019}
\bibfield{author}{\bibinfo{person}{David Salinas}, \bibinfo{person}{Valentin
  Flunkert}, {and} \bibinfo{person}{Jan Gasthaus}.}
  \bibinfo{year}{2019}\natexlab{}.
\newblock \bibinfo{title}{DeepAR: Probabilistic Forecasting with Autoregressive
  Recurrent Networks}.
\newblock
\newblock
\showeprint[arxiv]{1704.04110}~[cs.AI]


\bibitem[Shukla et~al\mbox{.}(2019)]%
        {dynamic_airline2019}
\bibfield{author}{\bibinfo{person}{Naman Shukla},
  \bibinfo{person}{Arinbj{\"{o}}rn Kolbeinsson}, \bibinfo{person}{Ken Otwell},
  \bibinfo{person}{Lavanya Marla}, {and} \bibinfo{person}{Kartik Yellepeddi}.}
  \bibinfo{year}{2019}\natexlab{}.
\newblock \showarticletitle{Dynamic Pricing for Airline Ancillaries with
  Customer Context}. In \bibinfo{booktitle}{\emph{Proceedings of the 25th {ACM}
  {SIGKDD} International Conference on Knowledge Discovery {\&} Data Mining,
  {KDD} 2019, Anchorage, AK, USA, August 4-8, 2019}},
  \bibfield{editor}{\bibinfo{person}{Ankur Teredesai}, \bibinfo{person}{Vipin
  Kumar}, \bibinfo{person}{Ying Li}, \bibinfo{person}{R{\'{o}}mer Rosales},
  \bibinfo{person}{Evimaria Terzi}, {and} \bibinfo{person}{George Karypis}}
  (Eds.). \bibinfo{publisher}{{ACM}}, \bibinfo{pages}{2174--2182}.
\newblock
\urldef\tempurl%
\url{https://doi.org/10.1145/3292500.3330746}
\showDOI{\tempurl}


\bibitem[Sutskever et~al\mbox{.}(2014)]%
        {sutskever2014sequence}
\bibfield{author}{\bibinfo{person}{Ilya Sutskever}, \bibinfo{person}{Oriol
  Vinyals}, {and} \bibinfo{person}{Quoc~V Le}.}
  \bibinfo{year}{2014}\natexlab{}.
\newblock \showarticletitle{Sequence to sequence learning with neural
  networks}.
\newblock \bibinfo{journal}{\emph{Advances in neural information processing
  systems}}  \bibinfo{volume}{27} (\bibinfo{year}{2014}).
\newblock


\bibitem[Tofallis(2015)]%
        {tofallis2015better}
\bibfield{author}{\bibinfo{person}{Chris Tofallis}.}
  \bibinfo{year}{2015}\natexlab{}.
\newblock \showarticletitle{A better measure of relative prediction accuracy
  for model selection and model estimation}.
\newblock \bibinfo{journal}{\emph{Journal of the Operational Research Society}}
  \bibinfo{volume}{66}, \bibinfo{number}{8} (\bibinfo{year}{2015}),
  \bibinfo{pages}{1352--1362}.
\newblock


\bibitem[Vaswani et~al\mbox{.}(2017)]%
        {vaswani2017attention}
\bibfield{author}{\bibinfo{person}{Ashish Vaswani}, \bibinfo{person}{Noam
  Shazeer}, \bibinfo{person}{Niki Parmar}, \bibinfo{person}{Jakob Uszkoreit},
  \bibinfo{person}{Llion Jones}, \bibinfo{person}{Aidan~N Gomez},
  \bibinfo{person}{{\L}ukasz Kaiser}, {and} \bibinfo{person}{Illia
  Polosukhin}.} \bibinfo{year}{2017}\natexlab{}.
\newblock \showarticletitle{Attention is all you need}.
\newblock \bibinfo{journal}{\emph{Advances in neural information processing
  systems}}  \bibinfo{volume}{30} (\bibinfo{year}{2017}).
\newblock


\bibitem[Warren(2017)]%
        {warren2017occupancy}
\bibfield{author}{\bibinfo{person}{Rex~Nelson Warren}.}
  \bibinfo{year}{2017}\natexlab{}.
\newblock \emph{\bibinfo{title}{Occupancy forecasting methods and the use of
  expert judgement in hotel revenue management}}.
\newblock \bibinfo{thesistype}{Ph.\,D. Dissertation}. \bibinfo{school}{Iowa
  State University}.
\newblock


\bibitem[Weatherford and Kimes(2003)]%
        {comparison2003}
\bibfield{author}{\bibinfo{person}{Larry~R. Weatherford} {and}
  \bibinfo{person}{Sheryl~E. Kimes}.} \bibinfo{year}{2003}\natexlab{}.
\newblock \showarticletitle{A comparison of forecasting methods for hotel
  revenue management}.
\newblock \bibinfo{journal}{\emph{International Journal of Forecasting}}
  \bibinfo{volume}{19}, \bibinfo{number}{3} (\bibinfo{year}{2003}),
  \bibinfo{pages}{401--415}.
\newblock
\showISSN{0169-2070}
\urldef\tempurl%
\url{https://doi.org/10.1016/S0169-2070(02)00011-0}
\showDOI{\tempurl}


\bibitem[Wooldridge(2008)]%
        {econometrics2008}
\bibfield{author}{\bibinfo{person}{J.M. Wooldridge}.}
  \bibinfo{year}{2008}\natexlab{}.
\newblock \bibinfo{booktitle}{\emph{Introductory Econometrics: A Modern
  Approach}}.
\newblock \bibinfo{publisher}{Cengage Learning}.
\newblock
\showISBNx{9780324581621}
\showLCCN{2007942361}
\urldef\tempurl%
\url{https://books.google.com/books?id=64vt5TDBNLwC}
\showURL{%
\tempurl}


\bibitem[Yan et~al\mbox{.}(2020)]%
        {traffic2020}
\bibfield{author}{\bibinfo{person}{Min Yan}, \bibinfo{person}{Junzheng Wang},
  \bibinfo{person}{Jing Li}, \bibinfo{person}{Ke Zhang}, {and}
  \bibinfo{person}{Zimu Yang}.} \bibinfo{year}{2020}\natexlab{}.
\newblock \showarticletitle{Traffic scene semantic segmentation using
  self-attention mechanism and bi-directional {GRU} to correlate context}.
\newblock \bibinfo{journal}{\emph{Neurocomputing}}  \bibinfo{volume}{386}
  (\bibinfo{year}{2020}), \bibinfo{pages}{293--304}.
\newblock
\urldef\tempurl%
\url{https://doi.org/10.1016/j.neucom.2019.12.007}
\showDOI{\tempurl}


\bibitem[Yang et~al\mbox{.}(2014)]%
        {yang2014matching}
\bibfield{author}{\bibinfo{person}{Jie Yang}, \bibinfo{person}{Guoshan Zhang},
  {and} \bibinfo{person}{Kai Ma}.} \bibinfo{year}{2014}\natexlab{}.
\newblock \showarticletitle{Matching supply with demand: A power control and
  real time pricing approach}.
\newblock \bibinfo{journal}{\emph{International Journal of Electrical Power \&
  Energy Systems}}  \bibinfo{volume}{61} (\bibinfo{year}{2014}),
  \bibinfo{pages}{111--117}.
\newblock


\bibitem[Ye et~al\mbox{.}(2018)]%
        {customized2018}
\bibfield{author}{\bibinfo{person}{Peng Ye}, \bibinfo{person}{Julian Qian},
  \bibinfo{person}{Jieying Chen}, \bibinfo{person}{Chen{-}Hung Wu},
  \bibinfo{person}{Yitong Zhou}, \bibinfo{person}{Spencer~De Mars},
  \bibinfo{person}{Frank Yang}, {and} \bibinfo{person}{Li Zhang}.}
  \bibinfo{year}{2018}\natexlab{}.
\newblock \showarticletitle{Customized Regression Model for Airbnb Dynamic
  Pricing}. In \bibinfo{booktitle}{\emph{Proceedings of the 24th {ACM} {SIGKDD}
  International Conference on Knowledge Discovery {\&} Data Mining, {KDD} 2018,
  London, UK, August 19-23, 2018}}, \bibfield{editor}{\bibinfo{person}{Yike
  Guo} {and} \bibinfo{person}{Faisal Farooq}} (Eds.).
  \bibinfo{publisher}{{ACM}}, \bibinfo{pages}{932--940}.
\newblock
\urldef\tempurl%
\url{https://doi.org/10.1145/3219819.3219830}
\showDOI{\tempurl}


\bibitem[Zhang et~al\mbox{.}(2019)]%
        {deeplearningHotel2019}
\bibfield{author}{\bibinfo{person}{Qing Zhang}, \bibinfo{person}{Liyuan Qiu},
  \bibinfo{person}{Huaiwen Wu}, \bibinfo{person}{Jinshan Wang}, {and}
  \bibinfo{person}{Hengliang Luo}.} \bibinfo{year}{2019}\natexlab{}.
\newblock \showarticletitle{Deep Learning Based Dynamic Pricing Model for Hotel
  Revenue Management}. In \bibinfo{booktitle}{\emph{2019 International
  Conference on Data Mining Workshops, {ICDM} Workshops 2019, Beijing, China,
  November 8-11, 2019}}, \bibfield{editor}{\bibinfo{person}{Panagiotis
  Papapetrou}, \bibinfo{person}{Xueqi Cheng}, {and} \bibinfo{person}{Qing He}}
  (Eds.). \bibinfo{publisher}{{IEEE}}, \bibinfo{pages}{370--375}.
\newblock
\urldef\tempurl%
\url{https://doi.org/10.1109/ICDMW.2019.00061}
\showDOI{\tempurl}


\end{thebibliography}
\end{document}